\documentclass[%
  pra, aps, physrev,
  showkeys,
  twocolumn,
  showpacs,
  superscriptaddress,
  amsmath,amssymb,
  10pt
]{revtex4-2}
\usepackage{array}
\usepackage{color}
\usepackage{soul}
\usepackage{graphicx}

\usepackage[inline]{enumitem} 

\begin{document}
\title{Hybrid quantum-classical unsupervised data clustering based on the Self-Organizing Feature Map}

\author{I.~D.~Lazarev}
\affiliation{Federal Research Center of Problems of Chemical Physics and Medicinal Chemistry RAS, Chernogolovka, Moscow Region, Russia, 142432}
\affiliation{Faculty of Fundamental Physical-Chemical Engineering, Lomonosov Moscow State University, GSP-1, Moscow, Russia 119991}

\author{Marek Narozniak}
\affiliation{New York University Shanghai, 567 West Yangsi Road, Shanghai, 200122, China.}
\affiliation{Department of Physics, New York University, New York, NY, 10003, USA.}
\affiliation{sqrtxx.com, Krakowiak\'ow 4, 20{-}255, Warsaw, Poland}

\author{Tim Byrnes}
\affiliation{New York University Shanghai; NYU-ECNU Institute of Physics at NYU Shanghai; Shanghai Frontiers Science Center of Artificial Intelligence and Deep Learning, 567 West Yangsi Road, Shanghai, 200126, China.}
\affiliation{State Key Laboratory of Precision Spectroscopy, School of Physical and Material Sciences, East China Normal University, Shanghai 200062, China}
\affiliation{Center for Quantum and Topological Systems (CQTS), NYUAD Research Institute, New York University Abu Dhabi, UAE.}
\affiliation{Department of Physics, New York University, New York, NY 10003, USA}

\author{A.~N.~Pyrkov}
\email{Email address:pyrkov@icp.ac.ru}
\affiliation{Federal Research Center of Problems of Chemical Physics and Medicinal Chemistry RAS, Chernogolovka, Moscow Region, Russia, 142432}

\date{\today}

\begin{abstract}
  Unsupervised machine learning is one of the main techniques employed in artificial intelligence.  We introduce an algorithm for quantum assisted unsupervised data clustering using the self-organizing feature map, a type of artificial neural network. The complexity of  our algorithm scales as $O(LN)$, in comparison to
the classical case which scales as $O(LMN)$,
where $N$ is number of samples, $M$ is number of randomly sampled cluster vectors, and $L$ is number of the shifts of cluster vectors.
  We perform a proof-of-concept experimental demonstration of one of the central components on the IBM quantum computer and show that it allows us to reduce the number of calculations in the number of clusters.  Our algorithm exhibits exponential decrease in the errors of the distance matrix with the number of runs of the algorithm.
\end{abstract}

\maketitle

\section{Introduction}
The combination of big data and artificial intelligence (AI) ---  dubbed the fourth industrial revolution --- has profoundly affected the modern economy in a plethora of different ways from robotics to agriculture \cite{Lecun2015, ghahramani2015,schwab2017,esteva2019, tyrsa2017}.
Contemporary artificial intelligence methods based on neural networks (NNs) also have the potential to enhance the role of novel analytical methods in science and engineering \cite{kaggle2014, radovic2018, butler2018, radovic2018,NobelPrizeChemistry,NobelPrizePhysics}.
Despite such groundbreaking achievements, interpretability of such NNs is very low due to the quantity of parameters in such neural networks rapidly increase.
Paradoxically, the intrinsic mechanism of how such neural networks work and why they are so powerful remains unknown (in many cases it is regarded as universal black box oracle).
Additionally, these black-box oracles demand substantial resources for their training and exhibit a propensity to overfit when data is scarce \cite{sevilla2022compute}, in contrast to the human ability to effectively learn from ``small'' data.


Meanwhile, quantum computers are a rapidly developing technology that promises improvement in algorithmic scaling for particular types of problems \cite{nielsen2010quantum,Cerezo2021,Bharti2022,byrnes2021quantum}.
In particular, there has been much interest recently in applying quantum computing techniques to machine learning (ML) problems \cite{dunjko2018, biamonte2017, schuld2014, carleo2019, Benedetti2019, Cerezo2022, Sajjan2022}. In comparison with classical ML, it was recently shown that Quantum Machine Learning (QML) models can be trained with fewer parameters in comparison with classical NNs \cite{abbasPowerQuantumNeural2021,caroGeneralizationQuantumMachine2022}.
The main focus of early works in QML was in obtaining a quantum speedup \cite{biamonte2017, schuld2014} by applying quantum approaches to solve linear algebra problems, such as the Fourier transform or solving systems of linear equations \cite{wiebe2012,harrow2009,childs2017}.
Quantum algorithms were developed for  linear regression, principal component analysis, support vector machine, K-means algorithm and others \cite{lloyd2013,lloyd2014,dunjko2016,paparo2014,rebentrost2014}.
More recently, there has been attention on developing quantum neural networks \cite{kamruzzaman2019, schuld2014b, jeswal2019, broughton2020}.
The interest in quantum neural networks was inspired by progress in experimental quantum computing when it became possible to use parametrized quantum circuits,
where the parameters behave much like the weights of a neural network \cite{lewenstein1994,Benedetti2019}.
In particular, quantum algorithms for training and evaluating feed forward neural networks were developed,
which are one of the most usable neural network models \cite{allcock2018, tacchino2019}. Quantum models for convolutional neural networks, which may be suitable for the problems of learning of quantum states were also proposed \cite{cong2019, liu2019}. Recent results connecting the classical Bayesian approach to deep learning allowed for the development of a new algorithm for Bayesian deep learning on quantum computers \cite{zhao2019,daiQuantumBayesianOptimization2023}. For the problem of classification which is closely connected to the problem of clusterization, a protocol for quantum classification via slow feature analysis based on the use of quantum Frobenius distance was proposed \cite{kerenidis2018}.
In Ref.  \cite{Liu2022}, a quantum capsule network together with an efficient quantum dynamic routing algorithm was proposed.
The quantum versions of transformer architecture and transformer of quantum states were developed \cite{Viteritti2023,cherratQuantumVisionTransformers2024,khatriQuixerQuantumTransformer2024,zhangTransformerQuantumState2023}, that provides possibilities for quantum natural language processing. Inspired by the classical counterpart, the quantum denoising diffusion probabilistic model (QuDDPM) was proposed to enable efficiently trainable generative learning of quantum data \cite{zhangGenerativeQuantumMachine2024}.
Furthermore, many other hybrid and quantum protocols inspired classical neural networks were developed  \cite{kolleQuantumDenoisingDiffusion2024,li2019, Heese2022,killoran2019,bondarenko2019,dunjko2017,nautrup2019,foesel2018, rebentrost2018,purushothaman1997,verdon2019,cherny2019,byrnes2013,mishra2019, pyrkov2019, vinci2019, lu2019,mohseni2023deep,mohseni2024deep}.

At the same time, non-NN based hybrid quantum classical algorithms have become a new direction of significant interest \cite{mcclean2016,arute2020,akshay2020}.
Such hybrid algorithms involve quantum circuits used for part of the execution and are typically trained in a classical learning loop.
In particular, the Quantum Approximate Optimization Algorithm (QAOA) was developed to find approximate solutions to combinatorial optimization problems \cite{farhi2014,farhi2016}
and designed for problems such as MAX-CUT and Grover's algorithm \cite{arute2020,akshay2020,wang2018,jiang2017,huang2019,wecker2016,pagano2019,byrnes2018,liao2021quadratic}.
Another example of a well-known hybrid quantum classical algorithm is the Variational Quantum Eigensolver (VQE) for applications in quantum simulations
\cite{kandala2017,aspuru-guzik2005,lanyon2010,peruzzo2014}.
Currently, it is believed that implementation of quantum neural networks and hybrid quantum classical algorithms can be the main test bed to achieve practical quantum supremacy on Noisy Intermediate Scale Quantum (NISQ) devices \cite{preskill2018,Bharti2022,Cerezo2021}.

In this paper, we develop a hybrid quantum-assisted self-organizing feature map (QASOFM)
and apply it to the data clustering problem in an unsupervised manner. The self-organizing feature map (SOFM) was first proposed by Kohonen \cite{kohonen1990,kohonen1996,kohonen1997} as a self-organizing unsupervised learning algorithm which produces feature maps similar to those occurring in the brain \cite{solan2001}. SOFMs are used in many areas \cite{vilibic2016, guido1998, doszkocs1990, jones2012,mori2019,corsello2017,zhu2018,chea2016}
and in comparison with many other NNs, they apply competitive learning and preserve the topological properties of the input space \cite{kiviluotoa1996}.
The SOFMs represent data in a fundamentally topological way that allows one to perform dimensionality reduction.
Once it is trained, the map can classify a vector from the input space by finding the node with the smallest distance metric. The idea of the QASOFM is based on the use of the Hamming distance as a distance metric for training the SOFM that allows, in the quantum case, to reduce the number of distance calculations in the number of clusters and thus to speed up the original classical protocol.
In order to make our protocol more appropriate for the currently available generation of NISQ quantum devices, we optimized the quantum circuit for realizing the Hamming distance by reducing the number of one-qubit operations. We then apply it to a toy example of clustering paper abstracts and give a proof-of-concept realization of the quantum assisted SOFM on the IBM quantum computer \cite{ibmq}
and compare it to the classical case.

\begin{figure}
    \includegraphics[width=0.95\columnwidth]{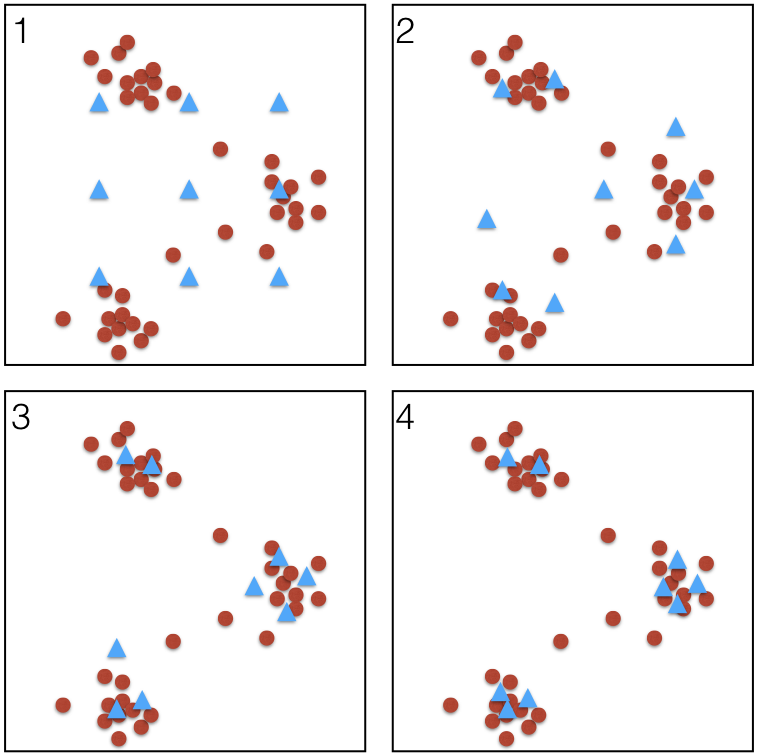}
  \caption{
    Schematic illustration of the clustering problem considered in this paper.
    Triangles represent clusters and circles are data points.
    The training process moves clusters to fit the data points.
    Note that there are fewer clusters than data points,
    which is the essence of dimensionality reduction,
    and is what permits the model to generalize the data.
  }
  \label{fig:sofm_fitting}
\end{figure}

\section{The quantum assisted self-organizing feature map}
\label{sec:qasofm}

\subsection{The classical algorithm}

The SOFM algorithm operates with a set of input objects, each represented by a $N$-dimensional vector,
and describes a mapping from a higher-dimensional input space to a lower-dimensional map space that preserves the topological properties of the input space, commonly referred to as a bi-dimensional map.  The input dimensions are associated with the features,
and the nodes in the grid (called cluster vectors) are assigned the $N$-dimensional vectors.
The components of these vectors are usually called weights.
Initially the weight components are chosen randomly.
The SOFM is then trained by adjusting the components through the learning process which occur in the two basic procedures of
selecting a winning cluster vector, also called the best matching unit (BMU), and updating its weights (Fig.~\ref{fig:sofm_fitting}).  The basic idea of the SOFM can be summarized by the four step process:
\begin{enumerate}
\item Select an input vector randomly from the set of all input vectors.
\item Find a cluster vector which is closest to the input vector.
\item Adjust the weights of the best matching unit and neurons close to it on the feature map in such a way   that these vectors  becomes closer to the input vector.
\item Repeat this process for many iterations until it converges.
\end{enumerate}

Specifically, when the BMU $\vec{w}_{c}$ for a input $\vec{x}(t)$ is selected,
the weights $\vec{w}_{i}$ of the BMU and its neighbors on the feature map are adjusted according to
\begin{equation}
    \label{eq:learning}
  \vec{w}_{i}(t + 1)
  = \vec{w}_{i}(t)
  + \theta(c, i, t) \alpha(t)
    \left(\vec{x}(t) - \vec{w}_{i}(t)\right) .
\end{equation}
Here, $ t $ is the iteration step,  $\alpha(t)$ is the monotonically decreasing learning rate and $\theta(c, i, t)$ is the neighborhood function (usually taken as a Gaussian or delta function), which defines the vicinity of the BMU labeled by an index $c$. The weights of the neighbors in the vicinity should also be adjusted in the same manner as for the BMU. This expression can be understood in the following way: if a component of the input vector $\vec{x}(t)$ is greater than the corresponding weight $ \vec{w}_{i}(t) $, increase the weight of the BMU and the weights indexed by $i$ and defined with the neighborhood function by a small amount defined by the learning rate $\alpha(t)$; if the input component is smaller than the weight, decrease the weight by a small amount. The larger the difference between the input component and the weight component, the larger the increment or decrement.

Intuitively, this procedure can be geometrically interpreted as iteratively moving the cluster vectors defined by the corresponding weight $ \vec{w}_{i}(t) $ (triangles in Fig.~\ref{fig:sofm_fitting}) in space one at a time in a way that ensures each move is following the current trends inferred from their distances to the input objects defined by $\vec{x}(t)$ (circles in Fig.~\ref{fig:sofm_fitting}).

In the original version of SOFM, the winning cluster vector is selected based on the Euclidean distance between an input vector and the cluster vectors. In this paper, we deal with a binary vector clustering problem, where we use the Hamming distance \cite{appiah2009, santana2017}.
Using a simple technique of encoding classical binary information into a quantum register \cite{trugenberger2001},
we introduce an optimized algorithm for calculating the matrix of Hamming distances between all input and cluster vectors at once.

\subsection{Optimized quantum scheme for Hamming distance calculation}
\label{subsec:qcircuit}

\begin{figure*}[t]
  \includegraphics[width=1.85\columnwidth]{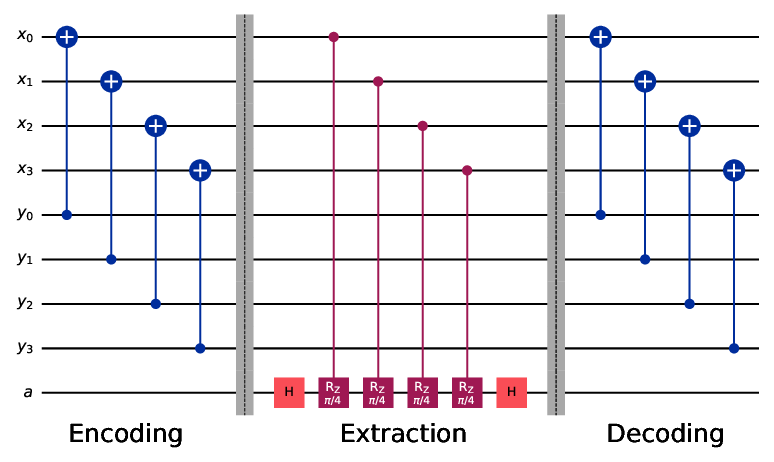}
  \caption{%
    The quantum circuit for the quantum parallelized Hamming distance calculation between all pairs of 4-bit binary vectors from two sets ${X}$ and ${Y}$.
    First, we encode the information about pairwise different qubits in a quantum state of the $X$-register by applying the CNOT gates.
    Second, the Hamming distance values are extracted to the amplitudes of superposition with the control phase rotation (see Eq.~(\ref{eq:controled_phase_rotation})) and Hadamard gates.
    Finally, the $X$-register is returned to the initial basis for information retrieval.   }
  \label{fig:qcircuit}
\end{figure*}

In Ref. \cite{trugenberger2001}, the concept of a probabilistic quantum memory was introduced. In this framework, quantum entanglement is utilized to significantly enhance the storage capacity of associative memories and to retrieve noisy or incomplete information. Specifically, the number of binary patterns that can be stored in such a quantum memory is exponential in relation to the number $ n $ of qubits, with a maximum capacity of $ p_{\text{max}} = 2^n $. This means it can optimally store all possible binary patterns that can be formed with $ n $ bits.

The retrieval process is probabilistic and involves postselection of the measurement results. This implies that the retrieval algorithm must be repeated until a certain threshold is met, or until the measurement of a control qubit yields a specific outcome. In the first scenario, the input is not recognized, while in the second, the output is determined by a probability distribution over the memory that is concentrated around the stored patterns that are closest to the input, as measured by Hamming distance. Here, we improve the approach for Hamming distance calculation proposed for realization of probabilistic quantum memory in Ref. \cite{trugenberger2001} and apply it to the realization of the QASOFM.

The overall procedure involves two registers of $n$ qubits each, denoted $\left| X \right\rangle$ and $\left| Y \right\rangle$, along with a single ancilla qubit $\left| a \right\rangle$.
Throughout the whole process, the $\left| Y \right\rangle$ register is used to store the cluster states.
At the beginning and end of the procedure, the $\left| X \right\rangle$ register stores the input vectors.
During the procedure it stores the differences between input vectors and cluster states.

Let us assume we have $N_x$ input vectors and $N_y$ cluster states.
The $i$th input vector and $j$th cluster vector are respectively denoted as $\left| x_i \right\rangle$, $\left| y_j \right\rangle$.
The registers $\left| X \right\rangle$ and $\left| Y \right\rangle$ are initialized to store the input vectors and cluster vectors according to
\begin{align}
    \label{eq:encodnig}
    \left| X \right\rangle  & = \frac{1}{\sqrt{N_x}} \sum\limits_{i=1}^{N_x} \left| x_i \right\rangle,  \\
    \left| Y \right\rangle&  = \frac{1}{\sqrt{N_y}} \sum\limits_{j=1}^{N_y} \left| y_j \right\rangle .
\end{align}
The two registers along with the ancilla qubit comprise the initial state of the quantum computer according to
\begin{equation}
    \label{eq:initial_state}
  \left| \psi_0 \right\rangle =
    \left| X \right\rangle
    \left| Y \right\rangle
    \left| 0 \right\rangle.
\end{equation}

Given this initial state we may begin the processing of the problem.  The quantum circuit that we follow is shown in Fig. \ref{fig:qcircuit}. We start by elementwise applying CNOT gates between all the qubits of the $X$ and $Y$ registers (the ``Encoding'' part of Fig. \ref{fig:qcircuit}), such that the state becomes
\begin{equation}
    | \psi_1 \rangle  =
    \frac{1}{\sqrt{N_x N_y}} \sum_{i, j=1}^{N_x,N_y}
    | d^{(1)}_{ij}, \dots, d^{(n)}_{ij} \rangle
    | y^{(1)}_j, \dots, y^{(n)}_j \rangle
    | 0 \rangle ,
\end{equation}
where $d^{(\alpha)}_{ij} = \mathrm{CNOT}(y^{(\alpha)}_i, x^{(\alpha)}_j)$, and $\alpha \in  [1,n]$  is the qubit index in the register.
At this stage of the computation the $\left| X \right\rangle$ no longer stores the input vectors,
instead it stores the information about pairwise difference of the qubits between the input vectors $\left| x_i \right\rangle$ and cluster vector $\left| y_j \right\rangle$.

The next step, depicted in Fig. \ref{fig:qcircuit} as the ``Extraction'' stage, is to extract the accumulated information about the differences between each pair $\left| x_i \right\rangle$ and $\left| y_j \right\rangle$ by projection onto the amplitude of the superposed state. This is achieved by applying the Hadamard gate on the ancilla qubit, followed by a controlled phase gate on $ | a \rangle  |d_{ij}^{(\alpha)} \rangle$, where the phase gate is defined by
\begin{equation}
    \label{eq:controled_phase_rotation}
    R_{(X,a)}(\phi) =
    \begin{pmatrix}
        1 & 0 & 0 & 0 \\
        0 & 1 & 0 & 0 \\
        0 & 0 & 1 & 0 \\
        0 & 0 & 0 & e^{-i\phi}
    \end{pmatrix} ,
    \quad \phi = \frac{\pi}{n}
\end{equation}
where $ | d_{ij}^{(\alpha)} \rangle $ is the control qubit and the ancilla qubit $\left| a \right\rangle$ is the target. Finally, another Hadamard gate is applied on the ancilla qubit (see Fig.~\ref{fig:qcircuit}).

After the first Hadamard on the ancilla qubit, the state is
\begin{equation}
    \left| \psi_2 \right\rangle =
    \frac{1}{\sqrt{N_x N_y}}\sum\limits_{i, j=1}^{N_x,N_y}
    \left| d_{ij} \right\rangle
    \left| y_j \right\rangle
    \dfrac{(\left| 0 \right\rangle + \left| 1 \right\rangle)}{\sqrt{2}},
\end{equation}
where $| d_{ij} \rangle = | d_{ij}^{(1)},\ldots,d_{ij}^{(n)} \rangle$.
Applying the controlled phase gate on $| x^{(\alpha)} a \rangle$, where $\alpha \in  [1,n]$  the state then becomes
\begin{multline}
    \left| \psi_3 \right\rangle = R_{(X,a)}\left(\dfrac{\pi}{n}\right)\left| \psi_2 \right\rangle
     = \dfrac{1}{\sqrt{2 N_x N_y}}
        \sum\limits_{i, j=1}^{N_x,N_y}
        \left| d_{ij} \right\rangle
        \left| y_j \right\rangle
        \left| 0 \right\rangle
        \\ + \dfrac{1}{\sqrt{2 N_x N_y}}
        \sum\limits_{i, j=1}^{N_x,N_y}
        \exp\left(\dfrac{-i \pi}{n}\sum\limits_{l=1}^n d^{(l)}_{ij} \right)
        \left| d_{ij} \right\rangle
        \left| y_j \right\rangle
        \left| 1 \right\rangle .
\end{multline}
Applying another Hadamard on the ancilla qubit we obtain
\begin{multline}
    \left| \psi_4 \right\rangle =
    \frac{1}{\sqrt{N_x N_y}}\sum\limits_{i, j=1}^{N_x,N_y}
    \exp \left(\dfrac{-i \pi}{2n}\sum\limits_{l=1}^n d^{(l)}_{ij} \right)
    \\ \times
        \left[ \cos\left(\dfrac{\pi}{2n}\sum\limits_{l=1}^n d^{(l)}_{ij} \right)
        \left| d_{ij} \right\rangle
        \left| y_j \right\rangle
        \left| 0 \right\rangle\right.
        \\+
        \left. i \sin\left(\dfrac{\pi}{2n}\sum\limits_{l=1}^n d^{(l)}_{ij} \right)
        \left| d_{ij} \right\rangle
        \left| y_j \right\rangle
        \left| 1 \right\rangle\right] .
\end{multline}
%

\begin{figure*}[t]
    \centering
  \includegraphics[width=1.85\columnwidth]{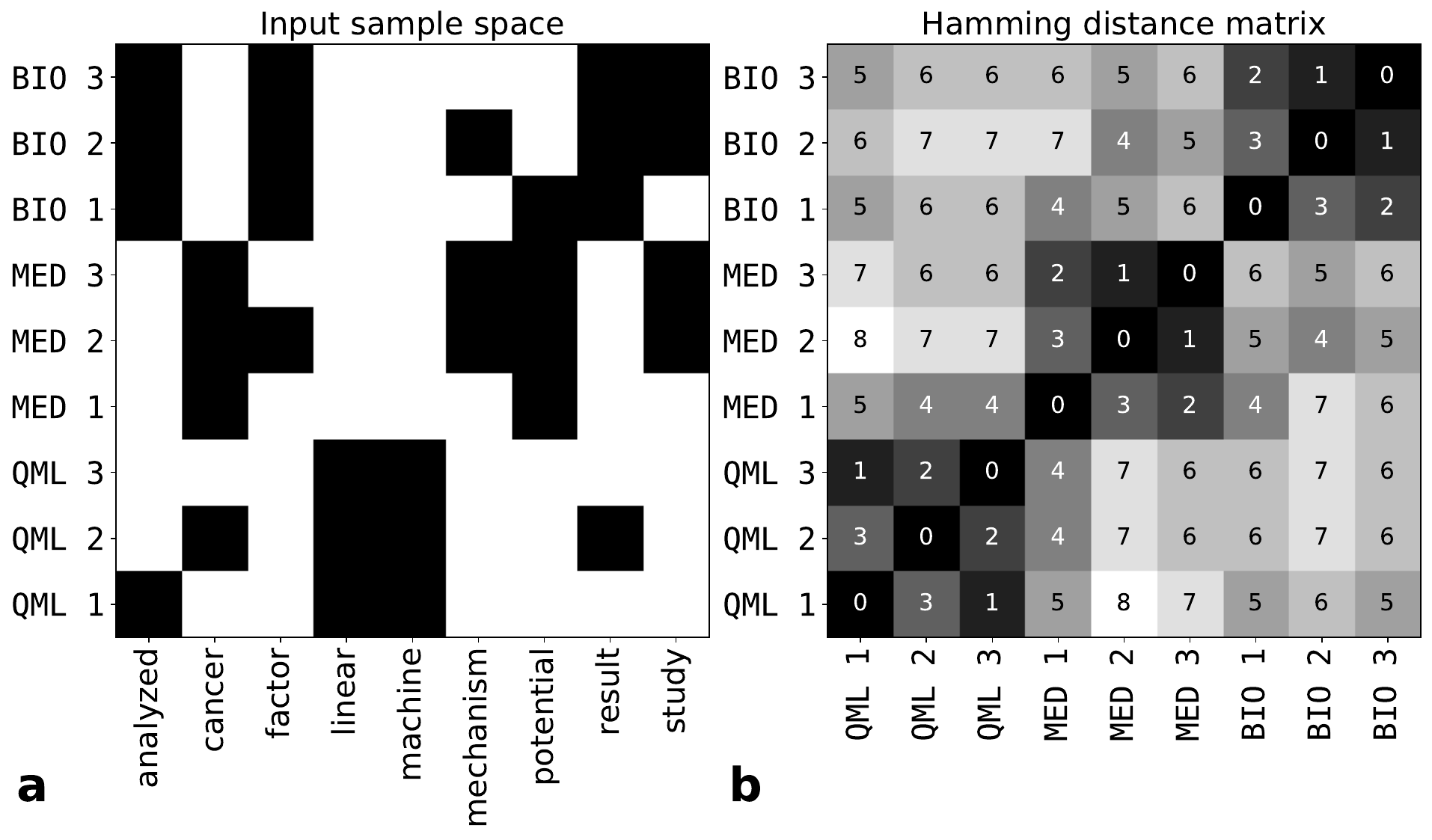}
  \caption{
    (a) Representation of the data set of abstracts with the bag-of-words \cite{weikang2016} model is shown.
    Each abstract is represented by a binary vector with 9 elements, corresponding to the 9 words on the horizontal axis.
    The samples are sorted into groups ``Quantum Machine Learning'' (QML), ``Cancer'' (MED) and ``Gene Expression'' (BIO) with 3 papers for each tag, for a total of 9 paper.
    (b) The Hamming distance between each vectorized abstract is shown as a number in the matrix.
  }
  \label{fig:vectorized_sample}
\end{figure*}

This completes the step of projecting differences between pairs of $\left| x_i \right\rangle$ and $\left| y_j \right\rangle$ onto the amplitude of the ancilla qubit.
The process is done in the $x$-basis, achieved by the surrounding Hadamard gates.
There are two possible measurement outcomes of the ancilla qubit.
Each pair of $\left|x_i \right\rangle$ and $\left| y_j \right\rangle$ forms a subspace of the Hilbert space,
the controlled phase gate ensures to change amplitudes of those outcomes within this subspace depending on how different the spin configurations between $\left| x_i \right\rangle$ and $\left| y_j \right\rangle$ are.

At this stage, the information regarding the differences between pairs of $\left| x_i \right\rangle$ and $\left| y_j \right\rangle$ is encoded in the amplitudes. In order to extract the Hamming distances between the relevant $\left| x_i \right\rangle$, $\left| y_j \right\rangle$ we return to our initial basis for register $\left| X \right\rangle$ by applying pairwise CNOT gates:
\begin{multline}
    \label{eq:final_state}
    \left| \psi_f \right\rangle =
    \mathrm{CNOT} (Y,X)\left| \psi_4 \right\rangle \\=
    \sum\limits_{i, j=1}^{N_x,N_y}
    \exp \left(\dfrac{-i \pi}{2n}\sum\limits_{l=1}^n d^{(l)}_{ij} \right)
        \left[ \cos\left(\dfrac{\pi}{2n}\sum\limits_{l=1}^n d^{(l)}_{ij} \right)
        \left| x_i \right\rangle
        \left| y_j \right\rangle
        \left| 0 \right\rangle\right.
        \\+
        \left. i \sin\left(\dfrac{\pi}{2n}\sum\limits_{l=1}^n d^{(l)}_{ij} \right)
        \left| x_i \right\rangle
        \left| y_j \right\rangle
        \left| 1 \right\rangle\right] .
\end{multline}
Thus we return into the initial basis and the amplitudes of the ancilla qubit are proportional to how different each pairs of $\left| x_i \right\rangle$ and $\left| y_j \right\rangle$ are.

From the statistics of the measurement outcomes of the final state~(\ref{eq:final_state}) we recreate the amplitudes of ancilla qubit states.
From those amplitudes estimations we are able to plot the distance matrix between two data sets of binary vectors.
The probability amplitude of the ancilla qubit outcomes captures the exact Hamming distance as the result of the preprocessing function.
There are two possible outcomes of measurement of the ancilla qubit, each has its probability amplitude and own interpretation of that amplitude.
For instance, for the $\left| 0 \right\rangle$ outcome, the larger the amplitude the smaller the Hamming distance,
and for the $\left| 1 \right\rangle$ outcome it is the other way around, the magnitude of the amplitude of that outcome is proportional to the Hamming distance.

Measuring the Hamming distance of a particular pair of input vectors $\left| x_i \right\rangle$ and cluster vectors $\left| y_j \right\rangle$ consists of extracting the relevant amplitude from the subspace that those states form,
this can be done using the following projection operator
\begin{align}
\Pi_{i,j} = &\left| x_i \rangle\langle x_i \right| \otimes \left| y_j \rangle\langle y_j \right| \otimes I .
\end{align}
Using the above projection operator, the subspace of the Hilbert space formed by a particular pair of input and cluster vectors can be traced out as
\begin{align}
    \rho_{i,j} &= \text{Tr}_{X,Y} (\Pi_{i,j} \left| \psi_f \rangle\langle \psi_f \right| \Pi_{i,j}) .
\end{align}
From the reduced density matrix, the following two amplitudes for the measurement results can be extracted
\begin{align}
    a_0(x_i,y_j) & = \frac{\left\langle 0 |\rho_{i,j}| 0 \right\rangle}{\text{Tr}(\rho_{i,j})}  \\
    a_1(x_i,y_j) & = \frac{\left\langle 1 |\rho_{i,j}| 1 \right\rangle}{\text{Tr}(\rho_{i,j})} .
\end{align}

Currently available quantum platforms are still subject to a substantial level of noise and extracting the exact distance from amplitude is still a difficult task. In order to reduce noise we average the measurement results over different states of the ancilla qubit, thus the measured Hamming distance between the input vector $\left| x_i \right\rangle$ and cluster vector $\left| y_j \right\rangle$ is
\begin{equation}\label{eq:amplitudes2distance}
    d_{i,j}^H \propto 1 - \frac{1}{2}(a_0(x_i,y_j) + (1-a_1(x_i,y_j))).
\end{equation}
The Hamming distance measured in this way is bounded as $0 \leq d_{i,j}^H \leq 1$,
where  $0$ indicates that $x_i$ and $y_j$ are identical and $1$ means they are the completely opposite in terms of their pairwise binary coordinates.

The number of controlled gate operations to define the full distance matrix matches the number of controlled gate operations in Ref. \cite{trugenberger2001}, where the original algorithm for calculation of the Hamming distance was introduced, but the number of remaining gates is reduced compared to Ref. \cite{trugenberger2001}, leading to a less deep circuit, which is significant for NISQ devices.


\section{Experimental demonstration}

We now show experimental results for a proof-of-concept demonstration of the algorithm introduced in the previous section with the use of the IBM quantum computer.  We perform unsupervised data clustering for three sets of paper abstracts from three different fields:
quantum physics \cite{qml0, qml1, qml2},
medicine \cite{med0, med1, med2}
and biology \cite{bio0, bio1, bio2}.
Each set consists of three papers selected at random that focus on one of the following topics:
``Quantum Machine Learning'',
``Cancer''
and ``Gene Expression''.
Abstracts were vectorized by the bag-of-words \cite{weikang2016} model in order to choose the most defining words in each data set (see Fig.~\ref{fig:vectorized_sample}) \cite{mctear2016}.
This model represents text as a multiset ``bag'' of its words taking into account only the multiplicity of words.
Preparing the bag-of-words we excluded the words that appear only in one abstract and more than in 4 abstracts and we also excluded the word ``level'' from consideration due to the frequent overlap between the clusters because it gives instabilities for both classical and quantum algorithms.
We restricted our bag-of-word size to 9 of the most frequent words from the full bags-of-word  due to limitations on the number of qubits.
After vectorizing and pre-processing  the data, the clusters are well-separated in terms of Hamming distance.
We observe that distances between the abstracts inside clusters are smaller than distances between the abstracts from different clusters,
showing successful self-organization (Fig.~\ref{fig:vectorized_sample}).

\begin{figure}[t]
  \includegraphics[width=\columnwidth]{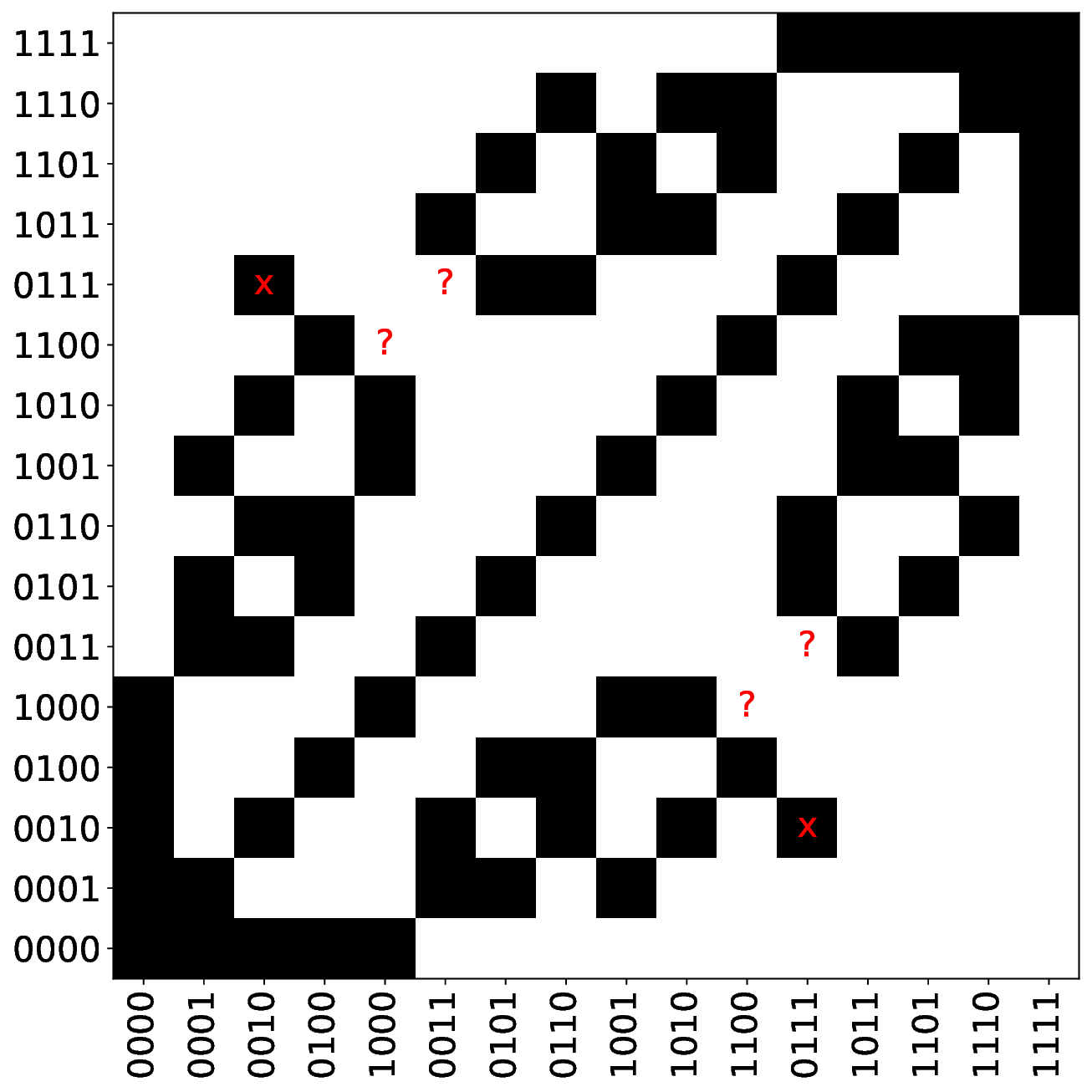}
  \caption{%
    Comparison of classical and quantum calculations of Humming distance matrix. Result of tomography of all vector pairs that differ not more than one bit --- nearest neighbors.  The distances are calculated as $\bar d_{ij} = (d_{ij} + d_{ji}) / 2$,
    where $d_{ij}$ was extracted from quantum state amplitudes with the use formula of ~(\ref{eq:amplitudes2distance}).
    The calculations of amplitudes were implemented on the IBM quantum backend ``ibm\_sherbrook'' with statistics of 8192 shots. Distance values less than 1.7 are marked black and coincide to the nearest neighbors.
    Question marks and crosses indicate missing and incorrect results compared to theoretical calculations.
    The accuracy compared to classical simulation is better than 97 $\%$.
  }
  \label{fig:distance_matrix}
\end{figure}

In order to check that our algorithm gives the expected results we compare it to classical calculations of the distance matrix on two data sets of binary vectors, as shown in Fig.~\ref{fig:distance_matrix}.
We see very good agreement between the distance matrices calculated classically and on the quantum computer.

\begin{figure*}[t]
  \includegraphics[width=1.85\columnwidth]{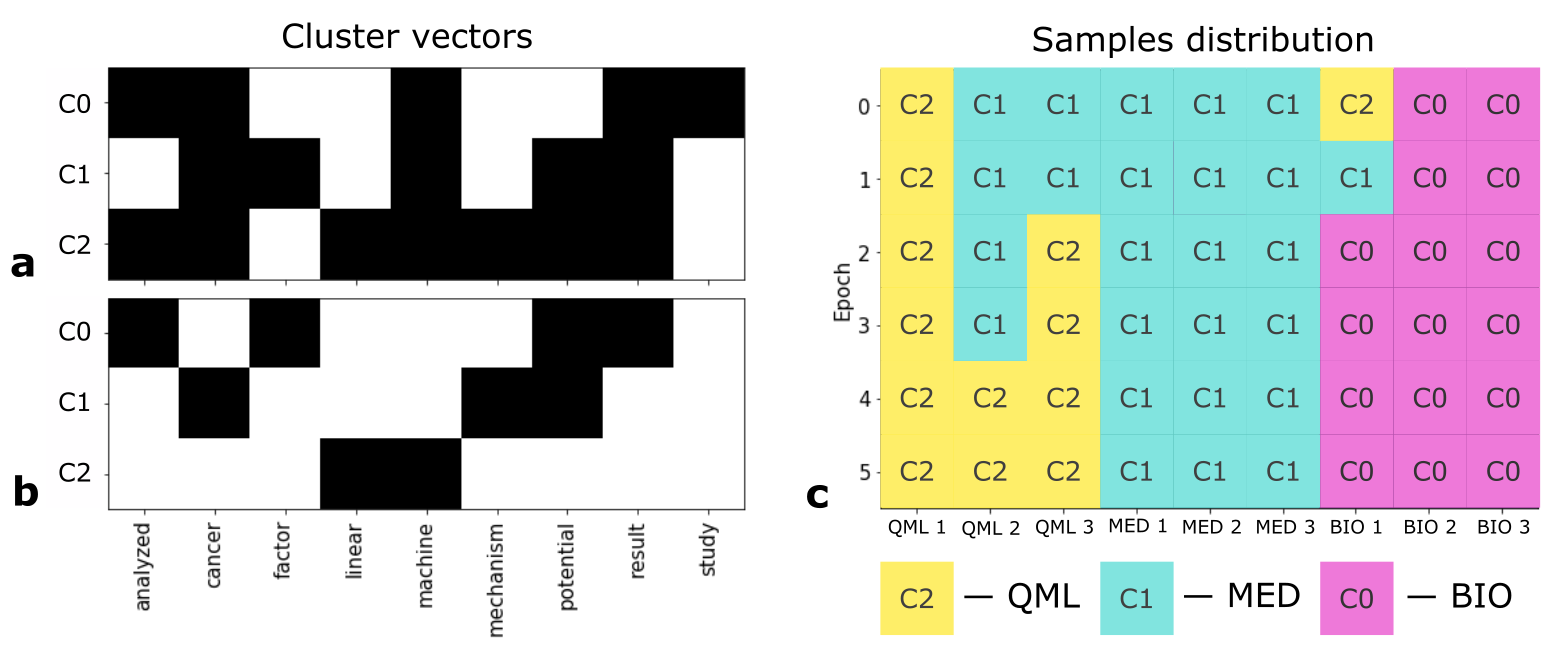}
  \caption{
    (a) Initial random binary vectors of cluster vectors (labeled as C0, C1, C2),
    for the selected 9 words from the bag of words model.
    (b) The result of applying our QASOFM implemented on the IBM Q Experience simulator backend.
    Vectors mean cluster elements for BIO, MED, QML groups,
    for the C0, C1, C2 cluster vectors, respectively.
    (c) The evolution of label distribution on each learning epoch.
  }
  \label{convergence}
\end{figure*}

An example of the QASOFM learning process is shown in Fig.~\ref{convergence}.
Initially, the cluster vectors were randomly chosen and the label of sample distribution is shown for the zeroth epoch in Fig.~\ref{convergence}(a). For the cluster  vectors (binary representation of rows of Fig. \ref{convergence}(a)) we use the simplest basis encoding into quantum  states with the standard initialization of QISKIT \cite{rathQuantumDataEncoding2024,qiskit} library.
Each epoch of the algorithm consists of a distance calculation between all data and cluster vectors
and requires 9 distance calculations in the  quantum implementation ($N$ in the general case)
or 27 distance calculations for the classical realization
($MN$ in the general case).
After the distance calculation from each sample to all cluster vectors at each epoch, we label each sample with the index of the closest cluster vector
and shift the closest cluster vectors to the sample vectors.
The shift is made by the change of the first binary element in the cluster vectors which does not match the binary element in the sample vector.
The evolution of the labels is presented in Fig.~\ref{convergence}(c).
Good convergence is already observed at the fourth epoch.

This proof of concept example shows that the development of NISQ hardware will allow to solve some practical problems in an unsupervised manner with a very simple encoding of categorical data to the quantum register.
Furthermore, it shows that it is possible to develop a distance based hybrid quantum classical algorithms
that speeds up classical counterparts and can be implemented on near-term quantum devices
developing distance based hybrid algorithms. Such algorithms,
that can learn in an unsupervised manner, for categorical data such as genomic data has significant interest.

\section{Complexity and scaling challenges}

In this section, we discuss the complexity and the challenges that are present for scaling QASOFM to large problems sizes.

\subsection{Complexity}

In classical SOFM \cite{kohonen1990}, first a distance calculation between a sample vector and all cluster vectors is made,
then the closest cluster vector is shifted towards the sample vector.
The complexity of algorithm, in the sense of the number of distance calculations, scales as $O(LMN)$,
where $N$ is number of samples,
$M$ is number of randomly sampled cluster vectors,
and $L$ is number of the shifts of cluster vectors.
In the QASOFM described in the previous section, distance calculations are realized on the quantum computer with the use of quantum circuit presented in Fig.~\ref{fig:qcircuit}. This approach allows one to reduce the number of operations in a number of cluster states with an optimized number of gates
that is possible to realize on quantum computers that are currently available.
The calculation of Hamming distance implemented step-by-step between each sample vector and all cluster vectors is realized in one operation. In this case, as there is only one input vector considered, the ``Decoding'' stage (Fig.~\ref{fig:qcircuit}) can be removed as the measurement no longer needs to indicate for which input vector the distance has been measured.
The complexity of the quantum assisted SOFM is then $O(LN)$.  Thus QASOFM algorithm has a quantum speedup by a factor of $ M $ over the classical algorithm.

\begin{figure*}[t]
  \includegraphics[width=1.85\columnwidth]{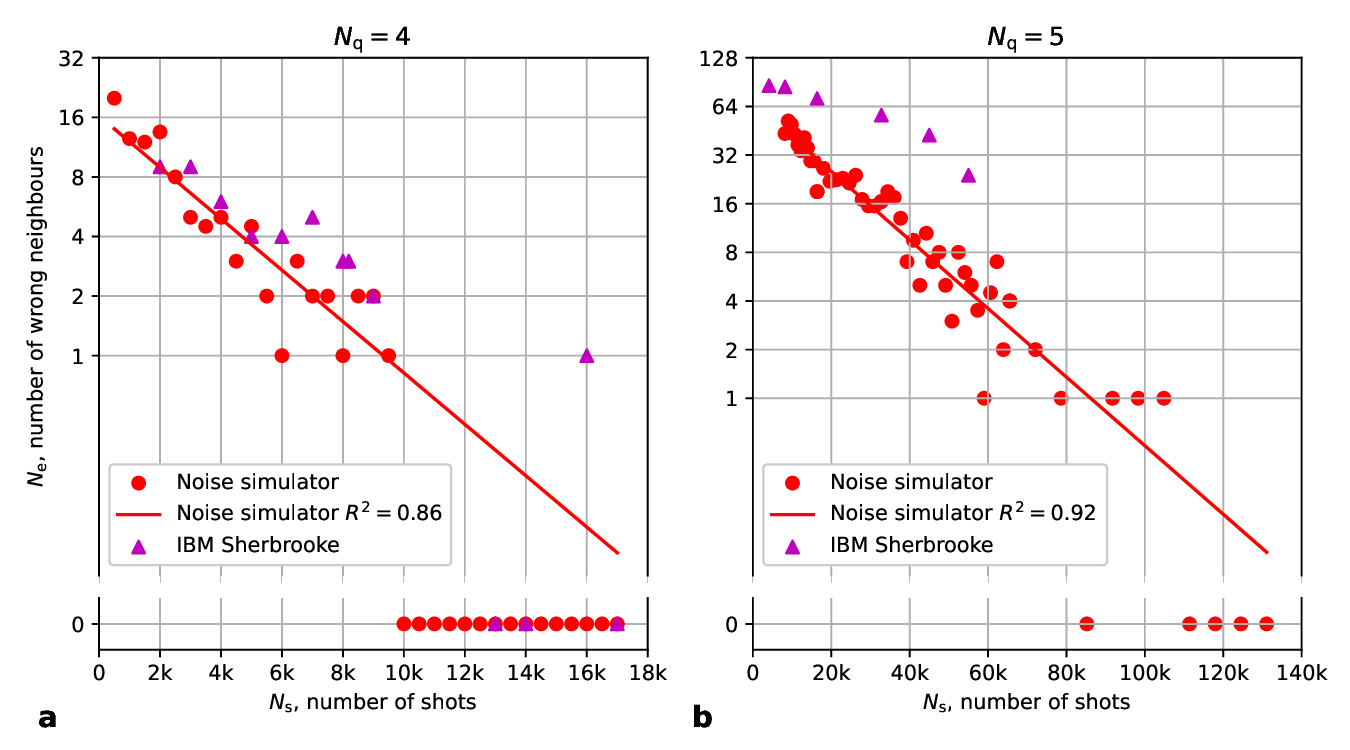}
  \caption{Error in the distance matrix of QASOFM versus number of runs of the quantum backend for (a) $4$ and (b) $5$ qubits. The triangles are the number of errors by the quantum scheme run on the IBM quantum backend named ``ibm\_sherbrook''.  The lines are the result of a regression analysis of the noise simulator data.  The coefficient of determination of the model is $R^2 = 0.86$ for the state space of $4$ qubits and $R^2 = 0.92$ for $5$ qubits.
  }   \label{fig:tomography-errors-by-shots-n4}
\end{figure*}

\subsection{Scaling challenges}

The above complexity estimate neglects the steps required to prepare the initial states (\ref{eq:encodnig}).  This is a standard problem that affects many quantum machine learning techniques and is called the ``input problem'' \cite{biamonte2017}.  It is typically assumed that there is an external resource (e.g. a qRAM) that can be used to prepare such states.  There are, however, potential ways to alleviate this by using a more effective encoding scheme into quantum states.  In the proposed QASOFM as given in Sec. \ref{sec:qasofm} we used the simplest encoding model. This model is inefficient with respect to quantum resources. Furthermore, the bag-of-word model is not optimal itself (we used it because of very small classical data needed to be embedded). The use of better word embedding models will allow to improve QASOFM performance. Also, the dimension of the input data is usually much larger than the number of cluster vectors. Consequently, the size of the subspace spanned on the cluster vectors is negligible compared to the whole space. This in theory provides opportunities to efficiently generate quantum circuits for initialization of such superposition of sparse quantum states with a smaller number of operations.

On the other hand, our QASOFM avoids the use exponential resources for the read-out  (i.e. no ``output problem'' due to the need of quantum tomography \cite{biamonte2017}).
The SOFM assumes that the cluster points do not have to move monotonically towards the cluster centroids,
and that the vectors closest to the true BMU have the highest probability of being measured.
Thus, the number of runs should only be chosen to measure distances to nearest neighbors with high accuracy, not all possible distances. The protocol implies finding all pairs of nearest vectors in the whole state space.
Fig.~\ref{fig:tomography-errors-by-shots-n4} shows the error versus number of runs for determination of the distance matrix for $4$  and $5$ qubits.
The points show the number of errors in the neighbors versus the the number of quantum circuit runs using a noise simulator using real parameters.  The number of errors decrease exponentially with the number of shots, which shows the efficient performance of our algorithm with respect to read-out.

Another challenge relates to the connectivity limitations of currently available quantum computers. In QASOFM, the ancilla qubit accumulates the distances in the phase information and as such it ideally requires a high connectivity to the $X $ register.  The topology of currently available superconducting quantum processors  presents a significant limitation to connectivity and consequently to the scaling of the QASOFM algorithm.  A possible solution is to create a large auxiliary network,
but it requires many two-qubit operations, which reduces the accuracy.
Another solution is to use platforms with a star architecture (or all to all connected) which are under development with trapped ion technology \cite{QuantinuumsHSeriesHits}.

\section{Discussions}
We have developed a quantum assisted SOFM and showed a proof-of-concept experimental demonstration
that it can be used to solve clustering problems in an unsupervised manner.
The procedure of solving such a clustering problem requires calculating the distance many times in iterative way,
which is calculated using a hybrid quantum-classical procedure.
We introduced an optimized circuit for Hamming distance calculations that can be implemented on currently available quantum computing devices with high fidelity.
Our quantum circuit performs the distance-computing component of a classical SOFMs algorithm in a parallelized way,
and in this way improves its performance.
The complexity of the quantum assisted SOFM scales as $O(LN)$
while the complexity of the fully classical SOFM scales as $O(LMN)$,
where $N$ is number of samples, $M$ is number of randomly sampled cluster vectors,
and $L$ is number of the shifts of cluster vectors.  We showed that the number of errors decreases exponentially with the number of runs of the algorithm, highlighting the efficiency of our algorithm. Due to wide use of classical SOFMs in different areas of modern research and technology,
this can give opportunities for the use of QASOFM in practical applications in the near term.

In addition, as our algorithm performs the Hamming distance calculation,
it has the potential to enhance any classical algorithm that relies on calculating distances between binary data. In particular, recently many algorithms are under development for quantized datasets with binary embeddings \cite{zhuangFastTrainingTripletBased2016,yiBinaryEmbeddingFundamental2015,AsymmetricDistancesBinary}. Furthermore, quantum protocols for Euclidean distance calculations have been developed \cite{yuQuantumAlgorithmsSimilarity2020,zardiniQuantumKnearestNeighbors2024}, which allows the generalization of QASOFM to non-binary data. In machine learning, data science, statistics and optimization, distance is a common way of representing similarity.  Calculating it between large data sets is common procedure
and our circuit could potentially enhance other distance based algorithms for binary data, particularly when the exact distance is not required,
and the knowledge of nearest vectors is sufficient.

\section*{Acknowledgments}
The work was partly supported by the
Russian Science Foundation (grant no 23-21-00507) and partly performed as a part of a state task, State Registration No. 124013000760-0.
T.B. is supported by the National Natural Science Foundation of China (62071301);
NYU-ECNU Institute of Physics at NYU Shanghai;
Shanghai Frontiers Science Center of Artificial Intelligence and Deep Learning;
the Joint Physics Research Institute Challenge Grant;
the Science and Technology Commission of Shanghai Municipality (19XD1423000,22ZR1444600);
the NYU Shanghai Boost Fund; the China Foreign Experts Program (G2021013002L);
the NYU Shanghai Major-Grants Seed Fund; Tamkeen under the NYU Abu Dhabi Research Institute grant CG008;
and the SMEC Scientific Research Innovation Project (2023ZKZD55).



\appendix

\section{Source code}

The source code is available on https://github.com/kephircheek/qasofm.




\bibliography{bibliography}

\end{document}